# Generation of electron beams from a laser-based advanced accelerator at Shanghai Jiao Tong University

Ahmed M. M. Elsied, Nasr A. M. Hafz,[a)] Song Li, Mohammad Mirzaie, Thomas Sokollik, Jie Zhang [b)]

Key Laboratory for Laser Plasmas (MOE) and Department of Physics & Astronomy, Shanghai Jiao Tong University, Shanghai 200240, China.

## Abstract

At Shanghai Jiao Tong University (SJTU) we have established a research laboratory for advanced acceleration research based on high-power lasers and plasma technologies. In a primary experiment based on the laser wakefield acceleration (LWFA) scheme, multi-hundred MeV electron beams having a reasonable quality are generated using 20–40 TW, 30 femtosecond laser pulses interacting independently with helium, neon, nitrogen and argon gas jet targets. The laser-plasma interaction conditions are optimized for stabilizing the electron beam generation from each type of gas. The electron beam pointing angle stability and divergence angle as well as the energy spectra from each gas jet are measured and compared.

Corresponding Author: Prof. Nasr A. M. Hafz

[b)] nasr@sjtu.edu.cn

[b)] jzhang1@sjtu.edu.cn

Department of Physics and Astronomy, Shanghai Jiao Tong University,

800 Dongchuan Road, Minhang District, Shanghai 200240, China

## Introduction

The field of laser-plasma electron acceleration has attracted a great attention since the first quasi-monoenergetic electron beam has been observed in 2004 [1-3]. In 2013 the generation of multiple GeV electron beams from two experiments was reported [4, 5]. This acceleration scheme was proposed by Tajima and Dawson in 1979 [6] and is called the "laser wakefield acceleration" (LWFA). The scheme works as follows: when an ultra-intense ultrashort laser pulse is focused in under-dense gaseous plasma its ponderomotive force repels the electrons sideward. However, the ions are practically immobile in this interaction due to ultra-short nature of the laser pulse. As the electrons are attracted back by the ions electrostatic field they overshoot around their initial positions, this sets up large amplitude electrostatic plasma wave in the wake of ultra-intense laser pulse [7, 8]. The phase velocity of the plasma wave equals to the group velocity of the laser pulse in the under-dense plasma which is almost the speed of light c. If the laser intensity is of the order of $10^{18}$ W/cm$^2$ or higher, and the plasma density is of the order of $10^{18}$ cm$^{-3}$, the wake's electrostatic field is of the order of 100 GV/m [9]. Therefore, the LWFA scheme has the potential to be the basis for a new technology which could lead to downsizing future particle accelerators. In order to reach the high intensity required to form the plasma wave, the laser pulse must be focused tightly, that in turn will shorten the interaction length, limited by diffraction of the Rayleigh length ($Z_R = \pi r_o^2 / \lambda$) [10]. numerous methods had been examined experimentally and by simulation to extend the propagation distance beyond the diffraction limit, most notably preformed channels [11, 12] and relativistically self-guided channels [13, 14]

The propagation of the laser through plasma causes a change in the index of refraction with the radius, at high laser power. This can be explained as follow, since the laser intensity changes with radius and the plasma frequency changes with the relativistic mass factor, the index of refraction which is given by $n = (1 - \omega_p^2 / \omega_o^2)^{1/2}$, will vary with radius [15]. Under these conditions, the plasma acts like a positive lens and focuses the beam. This effect has been found to have a power threshold given by $P_c[GW] = 16.5 n_c / n_e$, where n$_c$ is the critical density [16]. At laser power in order of 3P$_c$ the beam extends and forms a second focus. Further increase, multiple foci should occur, finally merge into a single channel [15].

In this paper we report on the generation of electron beams from the LWFA scheme in various gases, namely He, Ne, N, and Ar individually. We optimized the laser- plasma interaction conditions for the generation of electron beams of reasonable quality. We investigate the Ne gas for the first time as a target. However, He gas had been used for numerous previous experiments [17-19], the N gas was less tried [20] and the Ar is the least common gas [21] target in LWFA experiment. The aim is the generation of electron beams with low divergence and high-pointing stability. The current work could be useful to the ionization injection LWFA scheme [22-25] where a mixture of low-Z gas (e.g. He or H) and high-Z gas (e.g. N or Ar) are commonly used.

Therefore, the generation of electron beams and the study their parameters from each gas individually are meaningful.

**Experimental Setup**

A schematic diagram of the experimental setup is shown in Fig. 1. The research was carried out using a newly-installed Ti: Sapphire laser system at Key Laboratory for Laser Plasmas (LLP) at Shanghai Jiao Tong University in China. The laser system which is based on the CPA (chirped-pulse amplification) technique is operating at 10 Hz generating pulses with duration of 30 fs at the wavelength of 800 nm. The maximum peak power of the laser pulses are 200 TW. However, this experiment was the first using the facility and to avoid unexpected damages in the system, we only used 20–40 TW laser power. The laser pulses were focused onto the gas jet by using f/200 off-axis parabola (OAP). The focal spot size in vacuum was 30 μm (FWHM), and is shown in Fig. 2. The Rayleigh length based on ($1/e^2$) radius of the maximum intensity was 2.5 mm. Thus, the peak focused laser intensity and the corresponding normalized vector potential, $a_0$, were approximately 0.7–2.0×$10^{18}$ W/cm$^2$ and 0.6–1.0, respectively. The shot-to-shot laser energy was measured based on a calibration of the leaked laser energy measured by an energy meter, as shown in Fig. 1. The electron beam spatial profile on a DRZ phosphor screen was imaged onto a 14-bit charge-coupled device (CCD). The electron beam energy spectrum was measured by moving a dipole magnet into the beam path after the gas jet. The maximum magnetic field intensity was 0.94 T and the 2D-field intensity was mapped using a gauss meter. Then the magnetic field data points were used in a relativistic particle trajectory code (written using MATLAB) to calculate the electron beam energy. In the calculation, the beam's pointing angle (based on the data of Fig. 3, below) to the magnet entrance plane was taken into account. The laser-plasma interaction volume was imaged from the top by a 14-bit CCD. A narrow band-pass filter with transmission wavelength at ~ 500 nm was used in front of the CCD for collecting only the scattered laser light near the second harmonic wavelength.

**Experimental results and discussions**

A first step toward the observation of an electron beam from a laser-plasma accelerator is to make sure that the laser pulse has been guided during its propagation in the plasma, which means that one has to observe laser-plasma channels. Typical laser-plasma channels captured by the top-view CCD is shown in Fig. 3 where 2.3 mm ~ 3.3 mm long laser-plasma channels were observed. The laser-plasma channels are narrow over the full length; except for the Ar gas jet where the channel starts narrow then expands later on. We believe that such channel behavior in Ar might be a result of cluster formation which is common in the Ar gas jet expansion at atmospheric pressures [26]. In order for the laser-plasma accelerators to be used for applications, the electron beam pointing stability has to be dramatically enhanced [27]. Figure 4 shows the pointing angles of electron beams generated by laser-driven He, Ne, N and Ar gas jets, respectively. The results are shown for different plasma densities and laser energies. For the case of helium gas jet, Fig. 4 (a), the electron beam showed quite stable pointing over the plasma density range $6.7 \times 10^{18} - 1.0 \times 10^{19}$ cm$^{-3}$ in which the beam pointing deviation from the zero pointing was less than 10 mrad in 58% of the lase shots. However, outside the above density range, the beams had larger pointing angles. For the neon gas jet case and in 60% of the laser shots the electron beam had a deviation of 12 mrad over a wide plasma density range of $2 \times 10^{18} - 3.6 \times 10^{19}$ cm$^{-3}$ as shown in Fig. 4(b). For the nitrogen gas jet case, the highest (in this experiment) electron beam pointing stability has been observed; in 47% of the shots the electron beam had a deviation of only 5 mrad at the plasma density $\sim 1\times 10^{18}$ cm$^{-3}$. Moreover, in 73% of the laser shots the electron beam had deviations less than 10 mrad in the plasma density range of $9.5\times 10^{17}$–$1.5\times 10^{18}$ cm$^{-3}$, Fig. 4(c). Finally, in the argon gas jet case which showed the lowest reproducibility and the least electron beam quality (it will be presented below) over the scanned range of plasma densities, $4.0$–$6.0\times 10^{18}$ cm$^{-3}$. The beam deviation over this plasma density range from the zero pointing was less than 15mrad. A quick conclusion on the beam pointing from the mentioned gas jets is the following: there is a clear *trend* for more stable electron beam pointing with small pointing angle fluctuations at low plasma densities. As the density goes higher the electron beam pointing angles goes larger. Thus for real applications of LWFA beams we suggest the use of gas jets at as low densities as possible.

Figure 5 shows electron beam spatial profiles from the 4 different gas jet targets upon interaction with the laser. Generally, the electron beam brightness was several thousand counts using the 14-bit CCD for all gas jets expect the electron beams from the Ar gas jet which were relatively weak. Typically, the electron beam divergence (FWHM) was < 5 mrad, which is well-collimated high-quality. In Fig. 5 we show the laser energy and the plasma density for each beam profile. We noticed that the beam divergence is sensitive to the plasma density in each gas jet target, as shown in Fig. 6. In the He gas jet case (Fig. 6a), the beam divergence shows a clear trend of increasing from 4 mrad at low density to > 10 mrad at the density of $1.2 \times 10^{19}$ cm$^{-3}$. A similar trend has been observed for the beams from the laser-driven Ar gas jet (Fig. 6d). However, the beam divergence trend of the beams from Ne and H gas jets has been initially increasing with density then decreased again at high densities (Fig. 6a-c). Finally, we have observed the electron beam divergence angle versus the laser intensity, Fig. 7. Fig. 7 (a, c) show clearly that well-collimated beams were generated from the He and N gas jets at a moderate laser energy around 0.5-0.7 J (corresponding to 16 TW-24 TW). Then, there was a clear trend for a larger beam divergence at higher laser intensities. The trend was unclear for electron beams from the Ne and Ar gas jet cases; we believe that we need more data in future research to get a clearer conclusion about this issue.

Finally, we have recorded the electron beam energy spectra from the laser-driven 4 different gas jets, the results are shown in Fig. 8. Obtained using 23 TW (laser energy of 0.68 J) at the plasma density of $4.2 \times 10^{18}$ cm$^{-3}$, the electron beam energy spectrum (top-panel) from the He gas jet is showing a quasi mono-energetic peak at ~ 120 MeV, while the maximum energy extends up to ~ > 300 MeV. Electron beam from 27 TW laser-driven N gas jet is showing a wider energy spectrum (Fig. 8 2$^{nd}$ panel) which has quasimonoenergtic peaks at ~ 110 MeV and 170 MeV. Electron beam energy spectra from the Ne and Ar gas jets are even wider with maximum energies around 150 MeV. However, as shown on the CCD counts (which are proportional to the beam yield), the He electron beam yield is lower than all beams generated from the other gas jets at roughly the same laser power. The highest yield was obtained from the Ne and Ar gas jets (3$^{rd}$ and 4$^{th}$ panels of Fig. 8). The obtained energies and energy spectra of Fig. 8 are generally expected; it is known that the electron self-injection mechanism in laser-driven pure He gas jet generally generates quasimonoenertic spectrum, especially at low plasma density. However, for high-Z gas jets the injection into the wakewave is controlled by the ionization process which generally generates large energy spread (continuous spectrum) beams.

## Conclusions

We have experimentally generated electron beams from He, Ne, N, and Ar gas jet targets. The properties of the electron beams from each gas jet has been studied and compared. Electron beams from the N gas jet had the lowest divergence (2.9 mrad) and pointing angle (< 5 mrad).

However, some electrons from the He gas jet reached 300 MeV of energy. Higher yields are observed in Ne and Ar gas jets.

## Acknowledgment

We appreciate Feng Liu and X. L. Ge for their help on the laser, and the useful discussions with Prof. K. Nakajima. This work was supported by the 973 National Basic Research Program of China (Grant No. 2013CBA01504), and the Natural Science Foundation of China NSFC (Grants: 11121504, 11334013, 11175119, and 11374209).

[26] O. F. Hagena, Z. Phys. D – Atoms, Molecules and Clusters, 1990, **17:** 157.

[27] N. A. M. Hafz, T. J. Yu, S. K. Lee, T. M. Jeong, J. H. Sung, J. Lee, Applied Physics Express, 2010, **3:** 076401.**Figure Captions**

1. A schematic diagram of the experimental setup
2. The focused laser spot
3. Images of the laser-plasma channels observed in different gas jets. The laser energy, the length of each channel, the plasma density and the laser direction propagation are shown on each panel.
4. Electron beam pointing angle in the laser-driven He, Ne, N and Ar gas jet targets at different densities.
5. 2D and 3D spatial profiles of electron beams from laser-driven He, Ne, N, and Ar gas jet targets.
6. Illustrates the electron beam's divergence angle (FWHM) dependence on the plasma density for He, Ne, N, and Ar gas jets.
7. Illustrates the electron beam's divergence angle (FWHM) dependence on the laser energy for He, Ne, N, and Ar gas jets.
8. Electron beam energy spectra from laser-driven He, N, Ne, and Ar gas jets. The plasma density and laser energy are shown for each case.

Fig. 1.

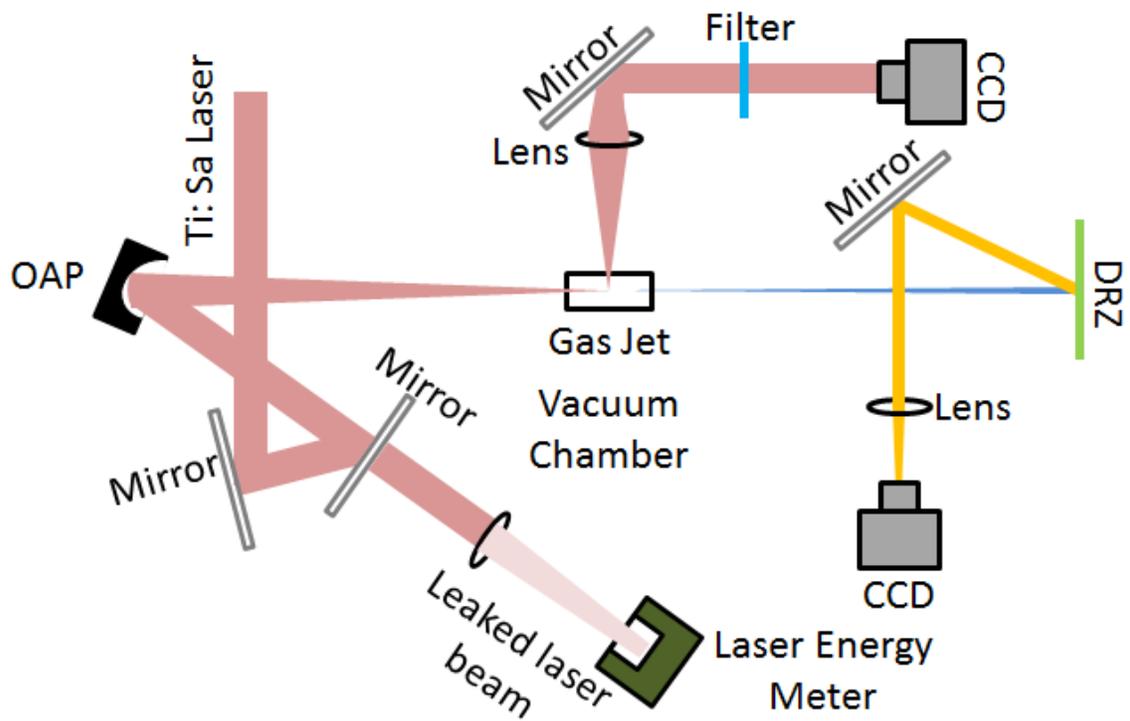

Fig. 2

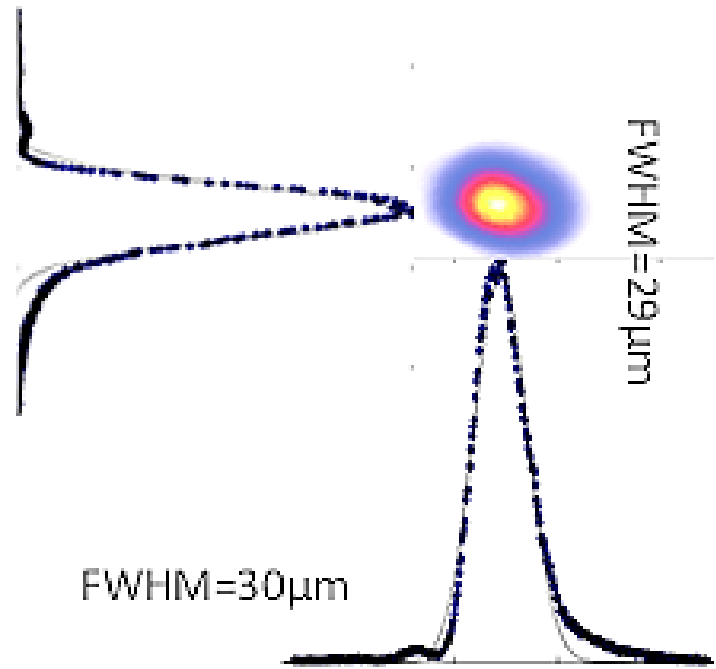

Fig. 3

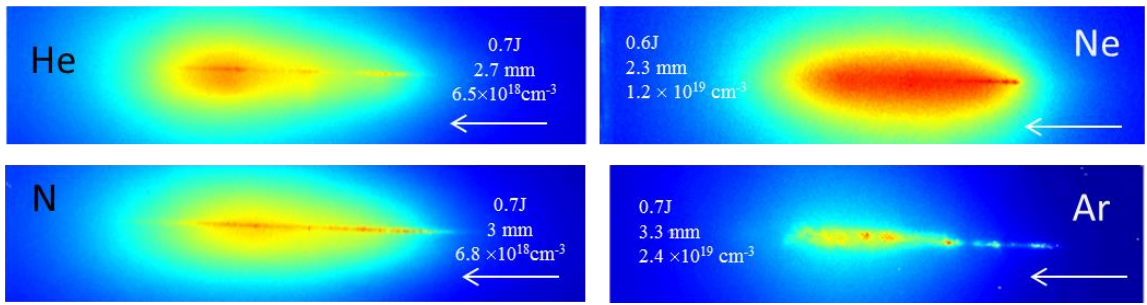

Fig. 4

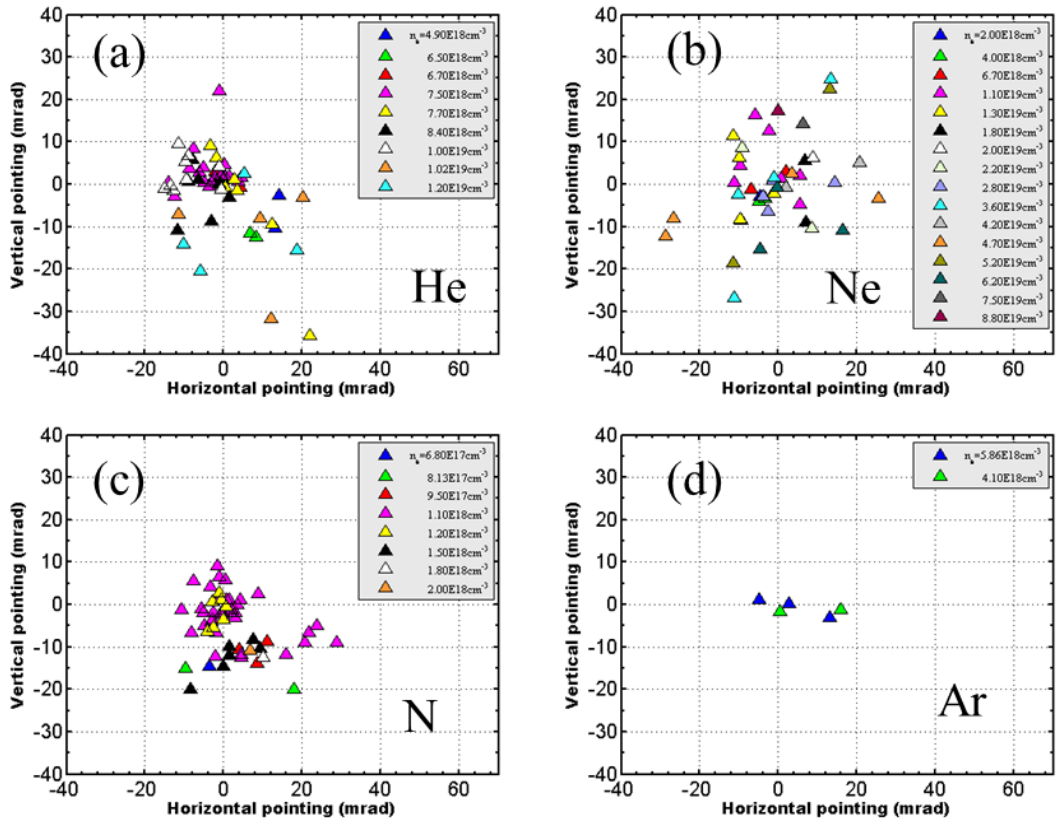

Fig. 5

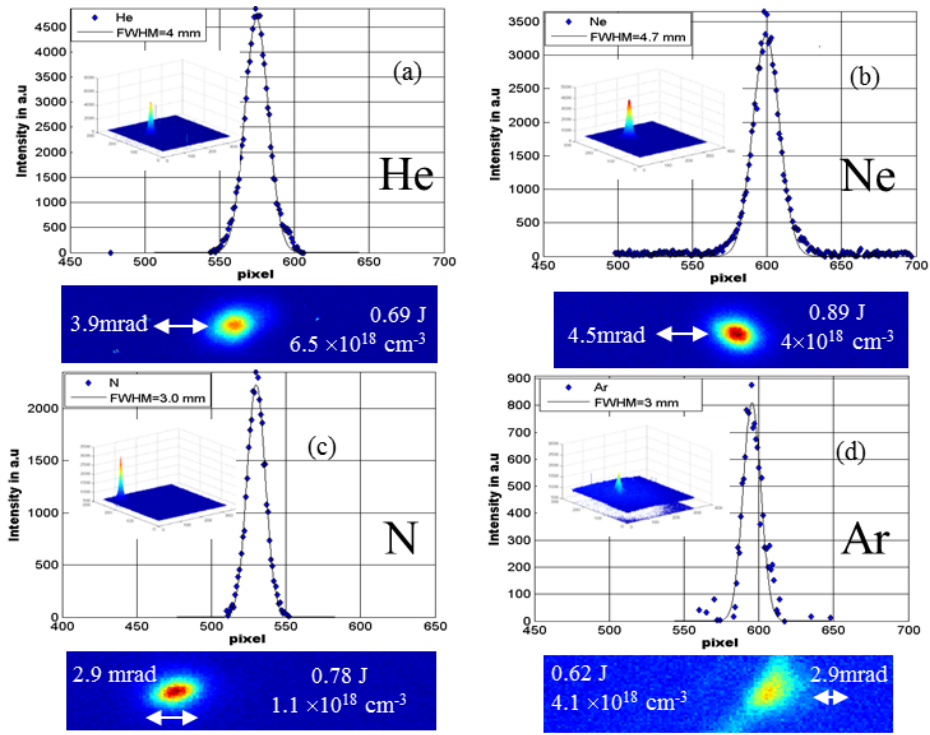

Fig. 6

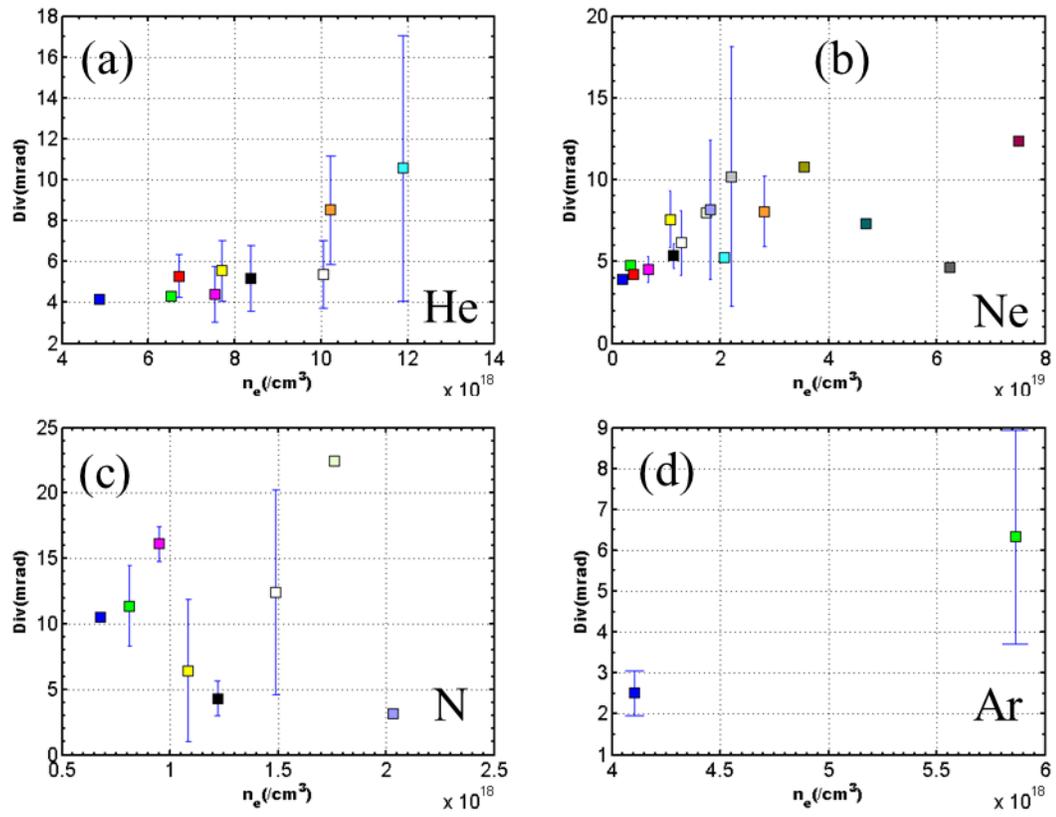

Fig. 7

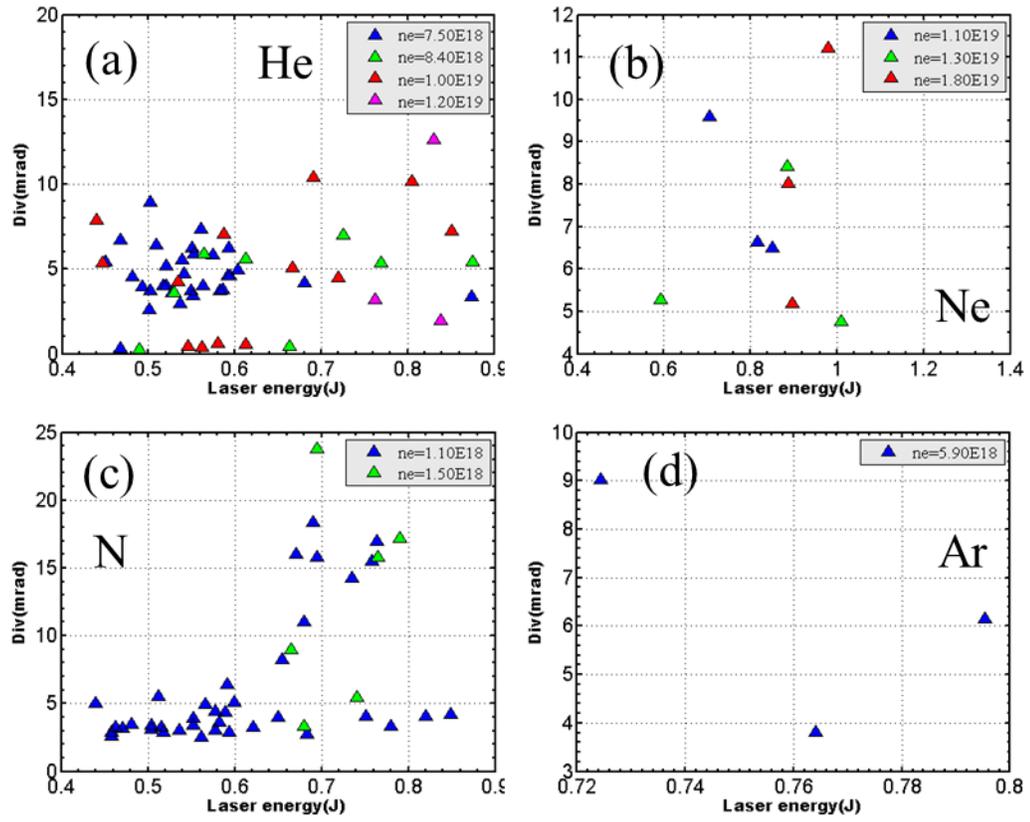

Fig. 8

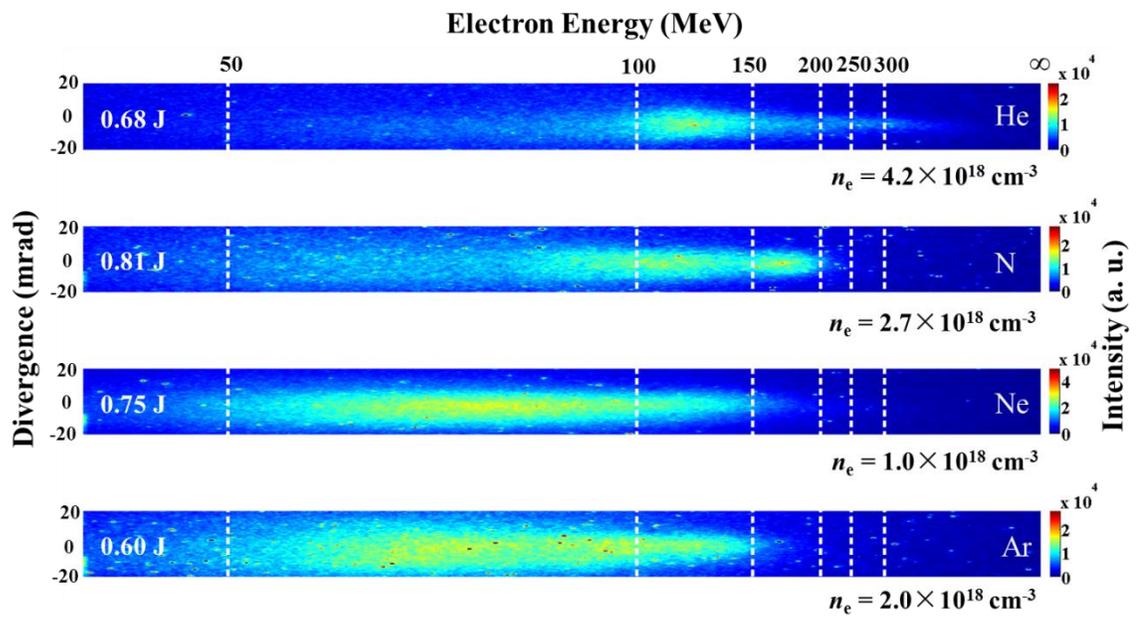